\documentclass[preprint,aps]{revtex4}

\usepackage{graphicx}% Include figure files
\usepackage{dcolumn}% Align table columns on decimal point
\usepackage{bm}% bold math
\usepackage{latexsym}
\usepackage{epsfig}
\usepackage{amsmath}
\usepackage{bm}% bold math

\newcommand{\be}{\begin{equation}}
\newcommand{\ee}{\end{equation}}
\newcommand{\beqq}{\setlength\arraycolsep{2pt}\begin{eqnarray}}
\newcommand{\eeqq}{\vspace{0cm} \end{eqnarray}}
\newcommand{\bea}{\begin{eqnarray}}
\newcommand{\eea}{\end{eqnarray}}

\begin{document}

\title{Can Dark Matter be a Scalar Field?}

\author{J. F. Jesus} \email{jfjesus@itapeva.unesp.br}
\affiliation{Universidade Estadual Paulista ``J\'{u}lio de Mesquita Filho'' -- Campus Experimental de Itapeva - R. Geraldo Alckmin, 519, Itapeva, SP, Brazil}

\author{S. H. Pereira} \email{shpereira@gmail.com}
\affiliation{Universidade Estadual Paulista ``J\'{u}lio de Mesquita Filho''\\ Departamento de F\'isica e Qu\'imica -- Campus de Guaratinguet\'a \\ Av. Dr. Ariberto Pereira da Cunha, 333\\
12516-410 -- Guaratinguet\'a, SP, Brazil}

\author{J. L. G. Malatrasi} \email{malatrasi440@gmail.com}
\affiliation{Universidade Estadual Paulista ``J\'{u}lio de Mesquita Filho'' - Campus Experimental de Itapeva - R. Geraldo Alckmin, 519, Itapeva, SP, Brazil}

\author{F. Andrade-Oliveira} \email{felipe.oliveira@port.ac.uk}
\affiliation{Institute of Cosmology and Gravitation, University of Portsmouth, Burnaby Road, PO1 3FX, Portsmouth, United Kingdom}
%\maketitle

%\pacs{04.30.Db, 09.62 +v, 98.80.Hw}
%\keywords{Dark energy, cosmic distance, Inhomogeneity Parameter}

%\bigskip
\begin{abstract}
In this paper we study a real scalar field as a possible candidate to explain the dark matter in the universe. In the context of a free scalar field with quadratic potential, we have used Union 2.1 SN Ia observational data jointly with a Planck prior over the dark matter density parameter to set a lower limit on the dark matter mass as $m\geq0.12H_0^{-1}$ eV ($c=\hbar=1$). For the recent value of the Hubble constant indicated by the Hubble Space Telescope, namely $H_0=73\pm1.8$ km s$^{-1}$Mpc$^{-1}$, this leads to $m\geq1.56\times10^{-33}$ eV at 99.7\% c.l. Such value is much smaller than  $m\sim 10^{-22}$ eV previously estimated for some models. Nevertheless, it is still in agreement with them once we have not found evidences for a upper limit on the scalar field dark matter mass from SN Ia analysis. In practice, it confirms free real scalar field as a viable candidate for dark matter in agreement with previous studies in the context of density perturbations, which include scalar field self interaction.
\end{abstract}

\maketitle

%%%%%%%%%%%%%%%%%%%%%%%%%%%%%%%%%%%%%%%%%%%%%%%%%%%%%%%%%%%%%%%%%%%%%%%%%%
\section{Introduction}

Nowadays the most accepted model in cosmology is known as $\Lambda$CDM model, where CDM stands for Cold Dark Matter and $\Lambda$ is the cosmological constant term, the latter being the main candidate to explain the current phase of acceleration of the universe and the former having a central role for the structure formation in the standard cosmology. The pressureless matter represents about 30\% of the total material content of the universe, where about 5\% is baryonic matter and 25\% is nonbaryonic dark matter (DM). The cosmological constant stands for the 70\% remaining part, sometimes also associated to a dark energy (DE) exotic fluid (see \cite{reviewDE} for a review).

The idea of an accelerating universe is indicated by type Ia Supernovae observations \cite{SN,ast05}, in agreement with $\Lambda$CDM, but the nature and origin of the dark matter is still a mystery (see \cite{DMrev,bookDM} for a review and \cite{tenpoint} for a ten-point test that a new particle has to pass in order to be considered a viable DM candidate).

The first candidate to DM as a scalar field is the axion, one of the solutions for the Charge-Parity problem in QCD \cite{kolb,weib}. The axion is essentially a scalar field with mass of about $10^{-5}$ eV, which has its origin at $10^{-30}$ seconds after the big bang. This candidate is until now one of the most accepted candidates to DM particles. Among others candidates for DM particles in the universe, the so called scalar field dark matter (SFDM) model \cite{sDM1,sDM2,sDM4,sDM4b,Magana12,bohua} is one of the most studied models in quantum field theory, and its applicability in cosmology to explain the different processes in the evolution of the universe has been investigated in the last decades (see \cite{scalarfield} for a review and references therein). In \cite{ringer} it is proposed a cosmological scalar field harmonic oscillator model, in agreement to $\Lambda$CDM model. Scalar fields are also used to drive the inflationary phase of the universe \cite{liddle,mukh}.  One of the first applications of a complex scalar field for structure formation of the universe was given by Press and Madsen \cite{press,mad}. Most recently, in \cite{sDM7} it was studied structure formation by assuming that dark matter can be described by a real scalar field. It was also investigated the symmetry breaking and possibly a phase transition of this scalar field in the early Universe. At low temperatures, scalar perturbations of SFDM leads to formation of gravitational structures. In \cite{sDM5} it has been shown that such model agree with rotation curves of dwarf galaxies and small and large low surface brightness (LSB) galaxies. The model was also extended to include the dark matter temperature and finite temperature corrections up to one-loop in the perturbative regime. Gravitational constraints imposed to dark matter halos in the context of finite temperature SFDM has been investigated in \cite{sDM6}. In \cite{sDM8} it was studied the Sextans dwarf spheroidal galaxy, embedded into a scalar field dark matter halo. In \cite{ccdm} the quantum creation of scalar particles was studied in the context of an alternative accelerating model of the universe.

It is important to notice that some models \cite{Magana12,bohua,sDM8} points to a mass of the scalar field of about $10^{-22}$ eV, while a recent work \cite{ringer} points to $10^{-32}$ eV. As we shall see, our model is in agreement to the last one.

In this work we study a real scalar field evolving in a Friedmann-Robertson-Walker background as a possible candidate to explain the dark matter in the universe. In section II, we develop the dynamics of the model and analyse the stability of the resulting system of equations. In Section III, we make an alternative derivation based on a change of variables, more suitable for numerical integration. In section IV, we constrain the free parameters of the model by using observational SN Ia data. We conclude in Section V and deduce the scalar field motion equation in the Appendix.

\section{The dynamics of the model}

We are interested in the study of the action of the form
\be\label{Stotal}
S = S^{grav} + S^{mat}\,,
\ee
where 
\be
S^{grav}=-{1\over 16 \pi G}\int d^4x \sqrt{-g}(R+2\Lambda)
\ee
is the standard Einstein-Hilbert action for the gravitational field, $R$ stands for the Ricci curvature scalar, $\Lambda$ is the cosmological constant parameter and $G$ is Newton's gravitational constant.

 The second term in (\ref{Stotal}) stands for the matter content of the universe, namely,

\begin{equation}\label{m63}
S^{mat}=\int d^4 x\sqrt{-g} \mathcal{L}^{sm} + \int d^4 x \sqrt{-g}\mathcal{L}^\phi\,,
\end{equation}
which is composed  of  the Lagrangian density of the standard matter contributions (radiation, baryons, neutrinos etc.), $\mathcal{L}^{sm}$, in addition to the Lagrangian density  of a real massive scalar field $\phi$ minimally coupled to gravity, $\mathcal{L}^\phi$, given by

\be
\mathcal{L}^\phi = {1\over 2}\partial^\mu \phi \partial_\mu \phi-V(\phi)\,,\label{lagrang}
\ee
where $V(\phi)$ is the potential.

Variation of the action $S$ with respect to $\phi$ leads to the equations of motion for the real scalar field. The equations of motion for the gravitational field are obtained by varying the action with respect to the metric $g^{\mu\nu}$, which leads to
\be
R_{\mu\nu}-{1\over 2}g_{\mu\nu}R=8\pi G T_{\mu\nu}\,,
\ee
where $T_{\mu \nu}$, the energy-momentum tensor of the matter fields, is defined as
\be
T_{\mu\nu}\equiv {2\over \sqrt{-g}}{\delta S^{mat}\over \delta g^{\mu\nu}}= T_{\mu\nu}^{sm}+T_{\mu\nu}^\phi\,.
\ee

Here we are assuming  that the standard matter term in the above equation can be approximated by a perfect fluid characterized by its energy density $\rho^{sm}$, pressure $p^{sm}$ and four-velocity $u^{sm}_\mu$. Therefore, we have
\be
T_{\mu\nu}^{sm}=(\rho^{sm}+p^{sm})u_\mu u_\nu-p^{sm}g_{\mu\nu}\,.\label{Tmat}
\ee

On the other hand, the scalar field part of the energy-momentum tensor is
\be
T^\phi_{\mu\nu} = \partial_\mu \phi \partial_{\nu}\phi-{1\over 2}g_{\mu\nu}\big[\partial^\alpha \phi \partial_{\alpha}\phi-V(\phi)\big]\,,
\ee
which can be rewritten in the form of a perfect fluid (\ref{Tmat}) using the definitions
\be
\rho_\phi \equiv {1\over 2} \partial^\alpha \phi \partial_{\alpha}\phi+V(\phi)\,,\hspace{1cm} p_\phi \equiv {1\over 2} \partial^\alpha \phi \partial_{\alpha}\phi-V(\phi)\,,\hspace{1cm} u_\mu \equiv (\partial_\mu\phi)/ \sqrt{\partial^\alpha \phi \partial_{\alpha}\phi}\,.
\ee

In particular, assuming that the field is homogeneous ($\partial_i \phi =0$), we have
\be
\rho_\phi={1\over 2}\dot{\phi}^2 + V(\phi)\,,\hspace{1cm} p_\phi ={1\over 2}\dot{\phi}^2 - V(\phi)\,.\label{rhop}
\ee

We restrict our treatment to a universe composed of a single baryonic component with energy density $\rho_b$, a cosmological constant term with energy density $\rho_\Lambda$, and the scalar field dark matter component $\rho_\phi$. The  equations of state of those components are respectively $p_b=0$, $p_\Lambda = - \rho_\Lambda$, and  $p_\phi=\omega_\phi \rho_\phi$,  where
\be
\omega_\phi = {p_\phi\over \rho_\phi}={{1\over 2}\dot{\phi}^2 - V(\phi)\over{1\over 2}\dot{\phi}^2 + V(\phi)}\,.\label{eqstate}
\ee

The Friedmann equations  obtained from this approach are (in units $c=\hbar=1$).
%\begin{subequations}
\begin{eqnarray}
\dot H&=&-\frac{\kappa^2}{2}\left(\dot{\phi}^2+\rho_{b}\right),\label{back1}\\
\ddot{\phi} &+& 3\,H \dot{\phi} + {dV\over d\phi}=0,\label{back2}\\
{\dot\rho_{b}}&+& 3\,H \rho_{b}=0, \label{back3}\\
{\dot\rho_{\Lambda}}&=&0, \label{back4}
\end{eqnarray}
%\end{subequations}
with the Friedmann constraint
\begin{equation}
H^2=\frac{\kappa^2}{3}\left(\rho_{\phi}+\rho_{b} + \rho_{\Lambda} \right),
\label{eq:FC}
\end{equation}
\noindent
where $\kappa^{2} \equiv 8\pi G$ and $H \equiv \dot{a}/a$ is the Hubble parameter. 

In order to study the system of equations (\ref{back1})-(\ref{back4}), we define the following dimensionless variables (accordingly to the variable transformations adopted in \cite{sDM4b} to a similar set of equations):
\begin{equation}
x\equiv \frac{\kappa}{\sqrt{6}}\frac{\dot{\phi}}{H},\quad
u \equiv \frac{\kappa}{\sqrt{3}}\frac{\sqrt{V}}{H},\quad
b\equiv \frac{\kappa}{\sqrt{3}}\frac{\sqrt{\rho_{b}}}{H},\quad
l \equiv \frac{\kappa}{\sqrt{3}}\frac{\sqrt{\rho_{\Lambda}}}{H}.
\label{eq:varb}
\end{equation}

Consequently, the Friedmann constraint (\ref{eq:FC}) reduces to
\be
x^2+u^2+b^2+l^2=1\,.\label{xubl}
\ee

Specializing to a quadratic scalar potential $V(\phi)=m^{2}\phi^{2}/2$, where $m$ is the mass of the scalar field, the background evolution of the universe is given by the equations (\ref{back1})-(\ref{back4}) as
%\begin{subequations}
\begin{eqnarray}
x'&=& -3\,x - s u+\frac{3}{2}\Pi\,x, \label{eq:dsb_x}\\
u'&=& s x +\frac{3}{2}\Pi\,u, \label{eq:dsb_u}\\
b'&=&\frac{3}{2}\left(\Pi -1\right)\,b, \label{eq:dsb_b}\\
l'&=&\frac{3}{2}\Pi\,l, \label{eq:dsb_l}\\
s'&=&{3\over 2}\Pi \,s, \label{eq:dsb_s}
\end{eqnarray}\label{eq:dsb}
%\end{subequations}
where a prime denotes derivatives with respect to the e-folding number $N=\ln a$,  
\be
s\equiv m/H\,\label{s} \,,
\ee
and
\begin{equation}
{3\over 2}\Pi\equiv -\frac{\dot H}{H^2}={3\over 2}(2x^2+b^2). 
\end{equation}

In order to study the stability of the system (\ref{eq:dsb_x})-(\ref{eq:dsb_s}) from a dynamical system approach \cite{booksystem}, we define the vector $\vec{v}=(x,\,u,\,b,\,l,\,s)$ and consider a linear perturbation of the form $\vec{v}\to \vec{v}+\delta \vec{v}$. Thus, the linearized system reduces to $\delta \vec{v'}=M \delta \vec{v}$, where $M$ is given by
\bea
M=\left(\begin{array}{ccccc}
-3+9x^2+{3\over 2}b^2 & -s & 3bx & 0 & -u \\
6xu+s & {3\over 2}b^2+3x^2 & 3bu & 0 & x \\
6xb & 0 & {9\over 2}b^2+3x^2-{3\over 2} & 0 & 0\\
6xl & 0 & 3bl & {3\over 2}b^2+3x^2 & 0\\
6xs & 0 & 3bs & 0 & {3\over 2}b^2+3x^2 
\end{array} \right)
\eea
and represents the Jacobian of $\vec{v'}$. 

The equilibrium points or fixed points $\{x_c,\, u_c,\, b_c,\, l_c,\, s_c\}$ of the phase space are:

(I) $\{\pm 1,\, 0,\, 0,\,0, \, 0\}$;

(II) $\{0,\, 0,\, \pm 1,\,0, \, 0\}$;

(III) $\{0,\, \pm \sqrt{1-l^2},\, 0,\, l, \, 0\}$;

(IV) $\{0,\, 0,\, 0,\, 1, \, s\}$.

According to the Friedmann constraint (\ref{xubl}), the case (I) represents a totally kinetic energy scalar dominated universe, the case (II) characterizes a totally baryonic dominated universe, the case (III) represents a mixture of scalar potential energy contribution and cosmological constant and the case (IV) a totally cosmological constant dominated universe. The case (IV) is the only one that admits $s>0$, leading to a mass $m>0$ for the dark matter particles from (\ref{s}). The eigenvalues of the matrix $M$ at the fixed points above are:

(I) $\{6,\, {3\over 2},\, 3,\, 3, \, 3\}$;

(II) $\{3,\, -{3\over 2},\, {3\over 2},\, {3\over 2}, \, {3\over 2}\}$;

(III) $\{0,\, 0,\, 0,\, -3, \, -{3\over 2}\}$;

(IV) $\{0,\, 0,\, -{3\over 2}+{1\over 2}\sqrt{9-4s^2},\, -{3\over 2}-{1\over 2}\sqrt{9-4s^2} , \, -{3\over 2}\}$.

As it is well known from the dynamical system approach \cite{booksystem}, when the real parts of the eigenvalues are positive the fixed point is unstable, and when the real parts are negative the fixed point is stable. When positive and negative real terms are present, the fixed point is a saddle point. Finally, when some of the eigenvalues are null, nothing can be said about the stability, and more accurate methods should be applied, as the center manifold theory. 

For the above set of eigenvalues, the most interesting one for us is the case (IV), where the stability is not completely discarded due to the null values of the two first eigenvalues. Beside that such case admits $s>0$, leading to a non-null mass to the dark matter particle. Although such case corresponds to a completely cosmological constant dominated universe at the fixed point, it is possible to estimate the mass as $m \sim H_0 \sim 10^{-33}$ eV if we assume $s\sim 1$. The case (III) is also interesting, since it also has some of the eigenvalues negative and others null, but it corresponds to $s=0$, leading to $m=0$. This case is also interesting since it admits the possibility of coexistence of cosmological constant and a potential part of $\phi$.

The limit of large mass can also be studied. From (\ref{eqstate}), we have $\omega_{\phi}\to -1$ when $m\to \infty$ or equivalently $V(\phi)>>\dot{\phi}^2/2$, which corresponds to $x<<u$. Taking the limit $x=0$, the fixed point of interest of the dynamical system is $\{0,\, 0,\, 0,\, 1, \, s\}$, indicating that the large mass limit corresponds to the $\Lambda$CDM model, with $l=1$ and $s\to \infty$.

In order to study the above system quantitatively we should try to integrate the dynamical system (\ref{eq:dsb_x})-(\ref{eq:dsb_s}). Nonetheless, we will do an alternative treatment in the next section, aiming the numerical integration.

\section{Alternative Derivation -- Change of variables}
\begin{figure}[t]
\centerline{ %\hspace{0.08\linewidth}
\epsfig{figure=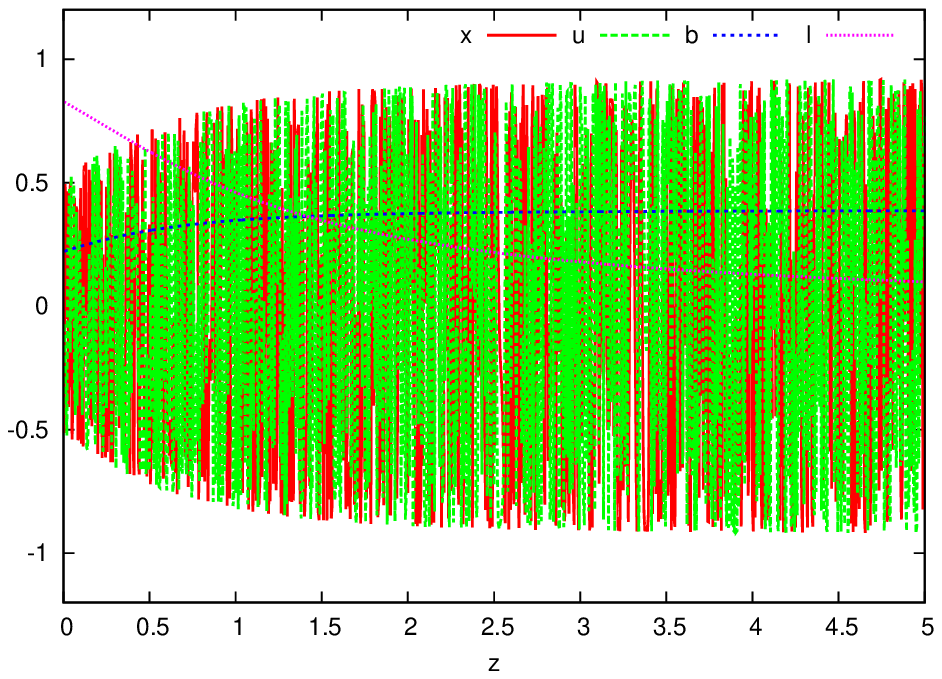,width=0.50\linewidth,angle=0}
\epsfig{figure=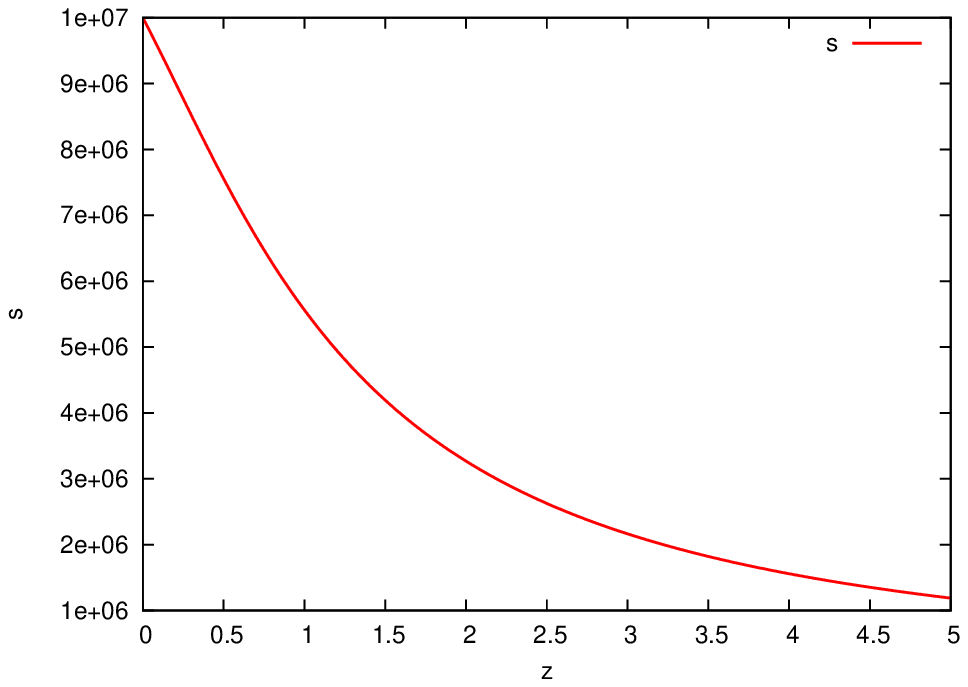,width=0.50\linewidth,angle=0}
}%-90}}height=3.8truein,
\caption{Panel {\bf a)} Evolution of variables $(x,u,b,l)$ in function of redshift for $\mu=10^7$ and $u_0=0$. Panel {\bf b)}. Evolution of variable $s$ in function of redshift for $\mu=10^7$ and $u_0=0$.}
\label{Evo_xubls}
\end{figure}

The system of differential equations derived in last section is useful for dynamical system analysis, as we had discussed in the last section. In order to constrain the SFDM mass we should solve these equations numerically. However, we must stress that this choice of variables may generate high frequency solutions in the variables $x$ and $u$, as already presented in Ref. \cite{Magana12} and shown in Fig. \ref{Evo_xubls}. For huge values of the mass $m$, these high frequency solutions require high computational efforts (as an illustration, one single evaluation for $m\sim10^{10}H_0^{-1}$ takes more than 24 hours with an Intel i5 processor), being desirable to avoid such parametrization.

Given that the behaviour of $\Omega_\phi$ and $\Omega_\phi=x^2+u^2$ are smooth  even for large values of the mass $m$, a reasonable choice of variables is $r\equiv\sqrt{\Omega_\phi}$, such that $r^2=x^2+u^2$, and the ``angle'' $\theta\equiv\tan^{-1}\frac{u}{x}$. Thus, the new variables are:
\begin{eqnarray}
 x=r\cos{\theta}\\
 y=r\sin{\theta}\,.
\end{eqnarray}
In some sense we are changing from cartesian to polar coordinates. The resulting system of equations is now:
\begin{eqnarray}
r'&=& -3r\cos^2(\theta) +\frac{3}{2}\Pi r, \label{eq:pol_r}\\
\theta'&=& s +\frac{3}{2}\sin^2(\theta), \label{eq:pol_q}\\
b'&=&\frac{3}{2}\left(\Pi -1\right)\,b, \label{eq:pol_b}\\
l'&=&\frac{3}{2}\Pi\,l, \label{eq:pol_l}\\
s'&=&{3\over 2}\Pi \,s \label{eq:pol_s}
\end{eqnarray}
where $\Pi=2r^2\cos^2(\theta)+b^2$. As can be seen, the equations are now even simpler on $r$ and $\theta$, and the system is invariant under the transformation $\theta\rightarrow\theta_1+\pi$, in such a way that it is periodic over $\theta$ with a period $\pi$. Thus, we seek to solve this system of equations over the vector of variables $\vec{r}=(r,\theta,b,l,s)$ with the corresponding initial conditions at $N=0$:
\begin{equation}
\vec{r}_0=\left(\sqrt{\Omega_{\phi0}},\theta_0,\sqrt{\Omega_{b0}},\sqrt{1-\Omega_{\phi0}-\Omega_{b0}},\mu\right)
\end{equation}
where $\theta_0$ becomes a free parameter, $\theta_0\in\left[-\frac{\pi}{2},\frac{\pi}{2}\right]$, we have used the normalization condition $\Omega_m+\Omega_\phi+\Omega_\Lambda=1$ and we have defined the parametrization of the mass:
\begin{equation}
 \mu\equiv\frac{m}{H_0}
\end{equation}

It is also  interesting to notice that in these new variables, the $\omega_\phi$ EOS assumes a nicely simple expression:
\begin{equation}
 \omega_\phi=\cos(2\theta)
\end{equation}

\begin{figure}[t]
\centerline{ %\hspace{0.08\linewidth}
\epsfig{figure=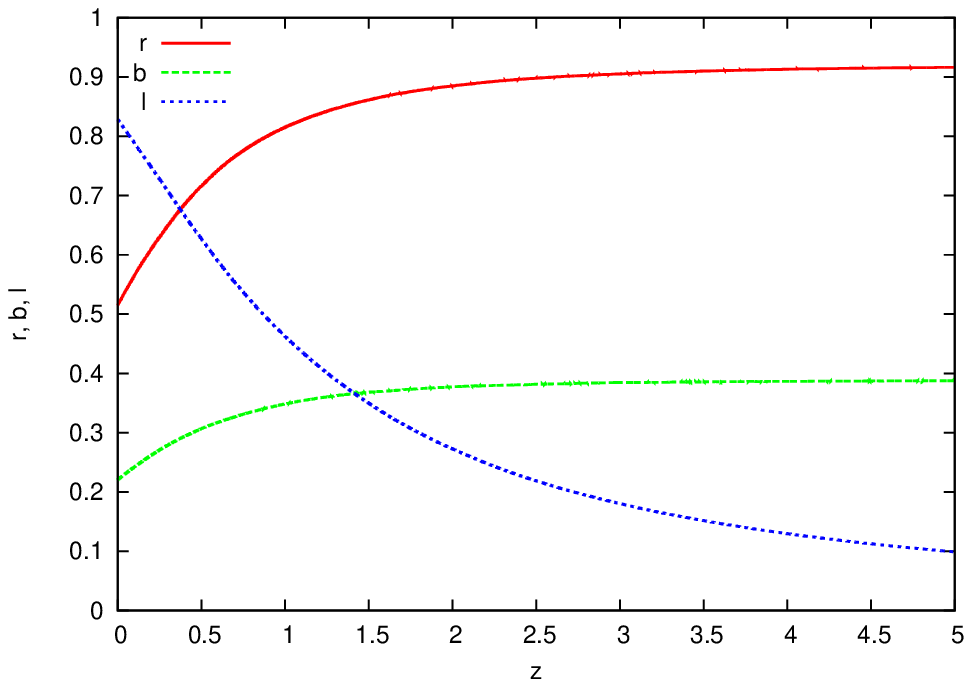,width=0.50\linewidth,angle=0}
\epsfig{figure=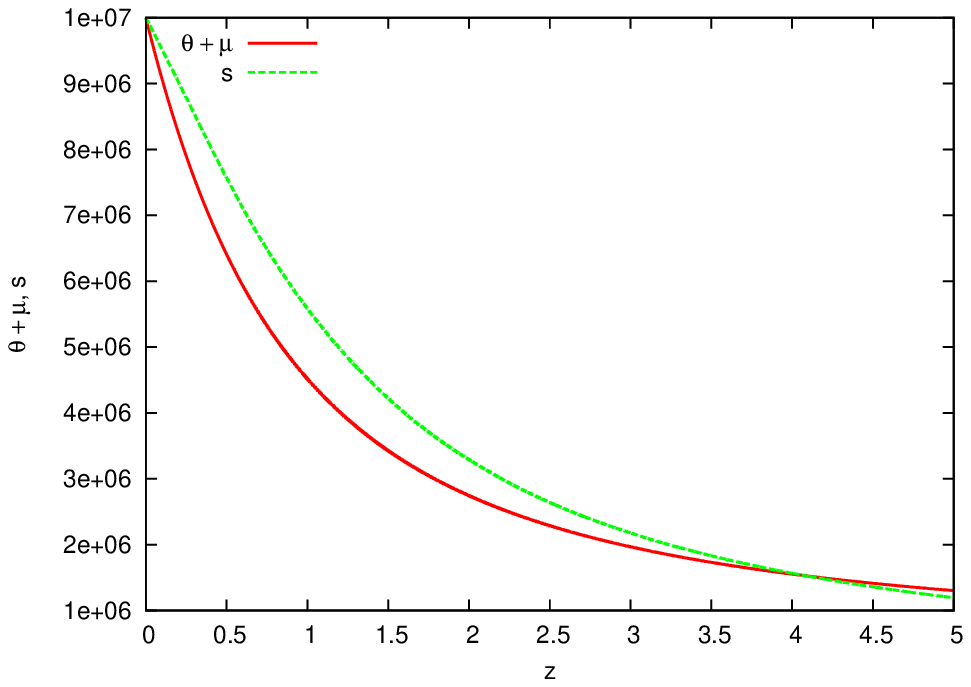,width=0.50\linewidth,angle=0}
}%-90}}height=3.8truein,
\caption{Panel {\bf a)} Evolution of variables $(r,b,l)$ in function of redshift for $\mu=10^7$ and $\theta_0=0$. Panel {\bf b)}. Evolution of variables $(s,\theta)$ in function of redshift for $\mu=10^7$ and $\theta_0=0$. We have plotted $\theta+\mu$ here in order for both variables stand on same axis scale.}
\label{Evo_rblqs}
\end{figure}

In Figure \ref{Evo_rblqs}, we may see that none of the ``polar'' variables present oscillations for higher mass values ($\mu=10^7$ on Fig. \ref{Evo_rblqs}).

\section{Constraining dark matter mass}
In order to obtain limits to the dark matter mass, we have to constrain the set of free parameters of the SFDM model, namely,  $\mathbf{s}=(\Omega_{\phi0},\theta_0,\Omega_{b0},\mu,H_0)$. Hence, even imposing spatial flatness, as predicted by inflation and indicated by CMB observations, we end up with 5 free parameters. Among these parameters, $\Omega_{b0}$ is the most well constrained to date, with compatible limits from BBN and CMB. We therefore choose to fix it at the value given by Planck and WMAP: $\Omega_{b0} = 0.049$ \cite{planck}, reducing the number of  free parameters to 4.

As a first attempt to constrain these free parameters we have considered the measurement of the Hubble parameter $H(z)$ in different redshifts. These kind of observational data are quite reliable because in general such observational data are independent of the background cosmological model, just relying on astrophysical assumptions. However, even using the current most complete compilation of $H(z)$ data, with 34 measurements \cite{sharov}, it was not enough to constrain the 4 free parameters, mainly because of the oscillatory behaviour of the model along with the relatively small number of degrees of freedom ($\nu=n-p=30$).

Next, we considered a SNe Ia data sample, which, although being more  dependent on the fiducial cosmological model, it consists of a large data sample and has passed recently through more refined and model independent methods of light curve fitting \cite{SuzukiEtAl12}.

The parameters dependent distance modulus for a supernova at a redshift $z$ can be computed through the expression
\be
\label{mudist}
\mu_{SN}(z|\vec{p}) = m_{SN} - M_{SN} = 5\log d_L + 25,
\ee
where $m_{SN}$ and $M_{SN}$ are respectively the apparent and absolute supernova magnitudes,  $\vec{p}\equiv (\Omega_{\phi0}, \theta_0, \mu,H_0)$ is the set of free parameters of the model and $d_L$ is the luminosity distance in units of Megaparsecs.

Since we have no analytic expression for $H(z)$, it is necessary to define $d_L$ through a differential equation. The luminosity distance $d_L$ can be written in terms of a dimensionless comoving distance $D$ by:
\begin{equation}
d_L=(1+z)\frac{H_0}{c}D\,.
\end{equation}

On its turn, the comoving distance can be related to $H(z)$, for a spatially flat Universe, by the following relation \cite{ClarksonEtAl08}:
\be
\label{clark}
\frac{dD}{dz}\equiv \frac{1}{E(z)},
\ee
where $E(z)\equiv\frac{H(z)}{H_0}$. Changing to the independent variable $N$ and given that $E=\frac{\mu}{s}$,  we arrive to:
%\begin{equation}
 %\frac{dD}{dN}=-\frac{e^{-N}}{E(N)}
%\end{equation}
%As $E=\frac{H}{H_0}$, then $E=\frac{\mu}{s}$, thus, we arrive at:
\begin{equation}
 D'(N)=-\frac{e^{-N}s}{\mu}
\end{equation}
which can be seen as a sixth equation to be solved simultaneously with the system (\ref{eq:pol_r})-(\ref{eq:pol_s}) with the initial condition for the comoving distance, $D(0)=0$. In order to constrain the free parameters of the model we considered the Union 2.1 SN Ia dataset from Suzuki et al. \cite{SuzukiEtAl12}. The best-fit set of parameters $\vec{p}$ was estimated from a $\chi^2$ statistics with %Amanullah =union 2.0
\be
\chi^2_{SN}=\sum^{n}_{i=1}\frac{\left[\mu_{SN}(z_i|\vec{p})-\mu_{SN,o,i}\right]^2}{\sigma_i^2}
\ee
where $\mu_{SN}(z_i|\vec{p})$ is given by (\ref{mudist}), $\mu_{SN,o,i}$ is the corrected distance modulus for a given SNe Ia at $z_i$ being $\sigma_i$ its corresponding individual uncertainty and $n=580$ for the Union 2.1 data compilation.

As usual on this analysis, we marginalize over the $H_0$ dependence by rewriting the distance modulus:
\begin{equation}
 \mu_{SN}(z)=5\log D_L(z)+M_*
\end{equation}
where $D_L=(1+z)D$ is dimensionless luminosity distance and $M_*\equiv25+5\log \frac{c}{H_0}$ comprises all the dependence over $H_0$. Then, we marginalize the likelihood over $M_*$:
\begin{equation}
 \tilde{\mathcal{L}}(\Omega_{\phi0}, \theta_0, \mu)=\int_{-\infty}^{+\infty} \mathcal{N}\exp\left[-\frac{1}{2}\chi^2(M_*, \Omega_{\phi0}, \theta_0, \mu)\right]dM_*
\end{equation}
where $\mathcal{N}$ is a normalization constant. The corresponding $\tilde{\chi}^2=-2\ln\left(\frac{\tilde{\mathcal{L}}}{\mathcal{N}}\right)$ is given by:
\begin{equation}
 \tilde{\chi}^2=C-\frac{B^2}{A}
\end{equation}
where $A=\sum_{i=1}^n\frac{1}{\sigma_i^2}$, $B=\sum_{i=1}^n\frac{5\log[D_L(z_i)]-\mu_{o,i}}{\sigma_i^2}$, $C=\sum_{i=1}^n\left\{\frac{5\log[D_L(z_i)]-\mu_{o,i}}{\sigma_i}\right\}^2$.

\begin{figure}[t]
\centerline{ %\hspace{0.08\linewidth}
\epsfig{figure=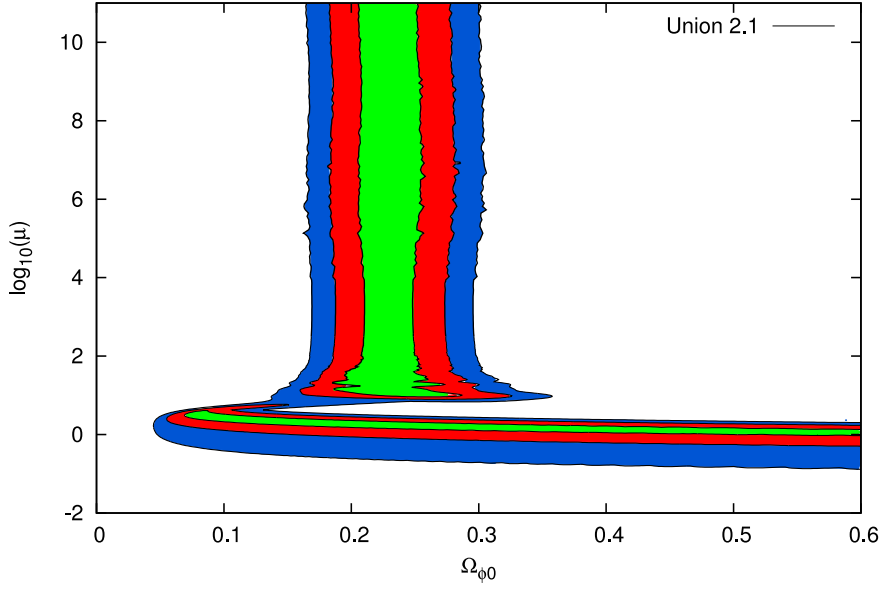,width=0.70\linewidth,angle=0}
}%-90}}height=3.8truein,
\caption{Confidence contours of SFDM from Union 2.1 SN data. The regions correspond to 68.3\%, 95.4\% and 99.7\% c.l. (green, red and blue regions, respectively).}
\label{contoursMuWphi}
\end{figure}

Since we are mainly interested on the constraints over the dark matter mass, we have  marginalized $\tilde{\mathcal{L}}$  over the parameter $\theta_0$. The result of this analysis can be seen in Fig. \ref{contoursMuWphi}, where we have plotted the statistical confidence contours on the plane $\log_{10}\mu$ - $\Omega_{\phi0}$.

As one may notice, there is a strong degeneracy for $\Omega_{\phi0}$ in the range of low DM mass ($10^{-1}\lesssim \mu \lesssim10$). For higher masses, $\mu\gtrsim10^2$, the degeneracy changes to prevent a mass determination.

In this case, restricting $\Omega_{\phi0}$ to the quite wide range $0.04<\Omega_{\phi0}<0.6$ (flat prior on this interval), we have found a $\chi^2_{min}=560.854$, which corresponds to $\chi^2_\nu=0.974$. We have found the best fit $\Omega_{\phi0}=0.22^{+0.38+0.38+0.38}_{-0.15-0.17-0.18}$ at $\Delta\chi^2=2.30$, $6.17$ and $11.83$, respectively. The best fit for $\mu$ was 19.95 and we may infer lower limits to $\mu$ as 0.94, 0.51 and 0.13 at the same $\chi^2$ levels.

\begin{figure}[t]
\centerline{ %\hspace{0.08\linewidth}
\epsfig{figure=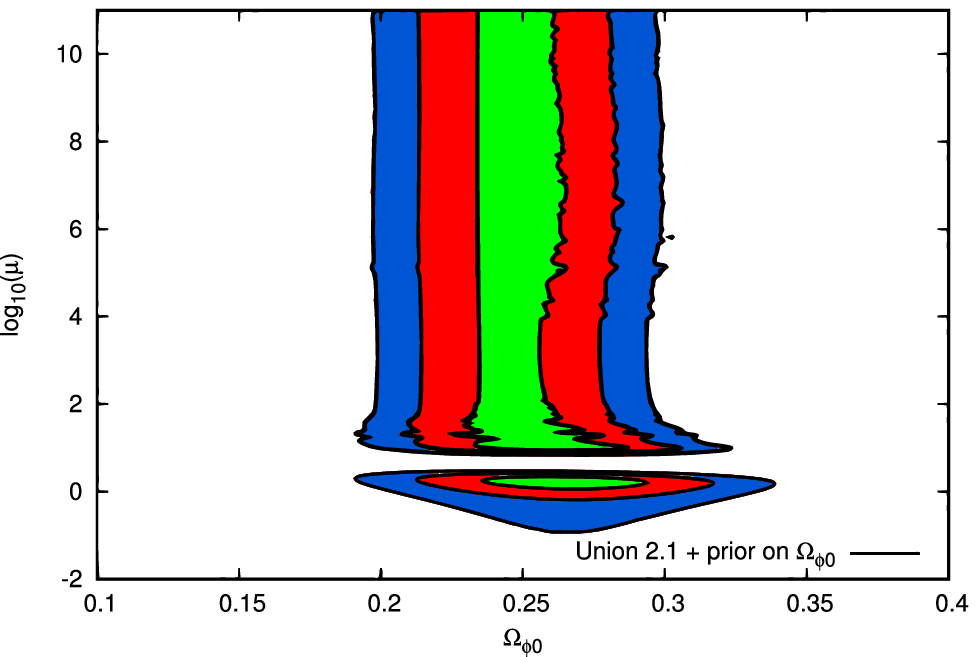,width=0.70\linewidth,angle=0}
}%-90}}height=3.8truein,
\caption{Confidence contours of SFDM from Union 2.1 SN data + Planck prior over $\Omega_{\phi0}$. The regions correspond to 68.3\%, 95.4\% and 99.7\% c.l. (green, red and blue regions, respectively).}
\label{contoursMuWphiPlPr}
\end{figure}

In order to alleviate the degeneracy over $\Omega_{\phi0}$, we use a prior over $\Omega_{\phi0}$ based on Planck results in order to constrain the mass $m$, or, at least, give it an inferior limit. As Planck (2013) \cite{planck} yields the limit $\Omega_{dm0}=0.265\pm0.011$, at 1$\sigma$, we use a Gaussian prior of $\Omega_{\phi0}=0.265\pm0.022$, which is approximately 2$\sigma$ of the Planck analysis. The result is shown on Figure \ref{contoursMuWphiPlPr}.

As one may see, $\Omega_{\phi0}$ is now better constrained in the interval $0.19<\Omega_{\phi0}<0.34$, at 3$\sigma$. We have found, from this analysis, $\chi^2_{min}=561.897$, $\chi^2_\nu=0.976$. The best fit is $\Omega_{\phi0}=0.252^{+0.041+0.065+0.086}_{-0.027-0.045-0.061}$. The degeneracy remains, however, over the dark matter mass, whose likelihood now presents two maxima, one for low mass and another for high mass. The best fit is $\mu=19.68$, with inferior limits given by 1.12, 0.64 and 0.12 at $\Delta\chi^2=2.30$, $6.17$ and $11.83$, respectively.

The lowest limit on this context, $\mu=0.12$, corresponds to a dark matter mass $m=0.12H_0^{-1}$, which corresponds to $m=(1.557\pm0.038)\times10^{-33}$ eV for a Hubble constant of $H_0=73\pm1.8$ km s$^{-1}$Mpc$^{-1}$, as indicated by the most recent analysis from the Hubble Space Telescope \cite{RiessEtAl16}.

%It must be noted, also, that Maga\~na and Matos have not made an statistical analysis of SFDM to find the DM mass estimate. Instead, they have preferred to make use of control parameter $k$ in order to have dynamical evolution similar to $\Lambda$CDM.

%\begin{figure}[t]
%\centerline{\epsfig{figure=sfdmHzData.eps,width=0.70\linewidth,angle=0}}%-90}}height=3.8truein,
%\caption{$H(z)$ data from Sharov and Vorontsova (2014) \cite{sharov} and some $H(z)$ curves with $h$ fixed on the best fit ($h=0.673$) and $\mu_{100}$ varying from the best fit, to 1$\sigma$ below and 1$\sigma$ above the best fit.}
%\label{sfdmEvo}
%\end{figure}

%%%%%%%%%%%%%%%%%%%%%%%%%%%%%%%%%%%%%%%%%%%%%%%%%%%%%%%%%%%%%%%%%%%%%%%%%%
\section{Concluding remarks}
In this work,  we investigated the hypothesis of the dark matter as a scalar field, where we have analysed the dynamics of a  quadratic potential for a single parameter free scalar field. Considering the SN Ia observational data, we have found a very low inferior limit to the dark matter mass,  namely, $m\geq 1.56\times10^{-33}$ eV. Although such lower limit is much smaller than the constraints obtained in \cite{Magana12,bohua,sDM8} (of about $10^{-22}$ eV), we have shown that any dark matter mass greater than this one is also compatible with SN Ia observations. Our constraints are  also much similar to the value recently obtained in \cite{ringer} ($m\sim 10^{-32}$ eV), which opens the question if the mass of the scalar field dark matter could be so small.

A very small mass may be in conflict with the structure formation of the universe, since  it consists of ultra hot dark matter. In the framework of structure formation, hot dark matter is disfavoured since in this case the galaxy-size density fluctuations would get washed out by free-streaming  leading to an earlier superclusters formation, while observations indicate early galaxy formation.

Experiments involving dark matter particles detections, e.g DAMA \cite{dama}, search for particles with mass in the range ($\sim$ 15 - 120 GeV). More recently \cite{bene}, an upper bound limit for dark matter mass of $\sim$ 197 TeV was established, based on relic abundance of thermal dark matter particles annihilating via a long-range interaction.

Studies of the density perturbation evolution should be made in order to obtain an upper limit on the scalar field dark matter mass. We may perform this analysis in a future work. Another forthcoming study is trying to test this result against different potential dependencies of the scalar field, including self-interaction which could eventually furnish an upper limit to dark matter mass.

\section{Appendix - The Motion Equation}
Another way to investigate the SFDM model is through the field motion equation. While it may be numerically intensive, this approach can be useful for dynamical interpretations. The motion equation for a scalar field is well known and can be obtained from the continuity equation as (\ref{back2}):
\begin{equation}
\ddot{\phi} + 3H \dot{\phi} + {dV\over d\phi}=0\,.
\end{equation}

Writing it in terms of the e-folds number, $N=\ln a$, which is more suitable for numerical integration, we have:
\begin{equation}
\phi'' + \left(\frac{H'}{H}+3\right)\phi' + \frac{V'(\phi)}{H^2}=0\,,
\label{eqphiN}
\end{equation}
where the primes at $\phi$ and $H$ denotes derivation with respect to $N$. Using the fact that $\rho_\phi$ can be written as $\rho_\phi(N)=\frac{1}{2}H^2\phi'(N)^2+V(\phi)$, we can write the Hubble parameter as
\begin{equation}
H^2=\frac{8\pi G\left[V(\phi)+\rho_{b0}e^{-3N}+\rho_\Lambda\right]}{3-4\pi G\phi'(N)^2}\,,
\end{equation}
where $\rho_\Lambda=$constant. Inserting this result into the motion equation, we arrive at an equation involving only field derivatives. However, in order to simplify the equations, we apply the free scalar field potential $\left(V(\phi)=\frac{1}{2}m^2\phi^2\right)$ and define the dimensionless quantities
\begin{eqnarray}
\Phi&\equiv&\sqrt{\frac{8\pi G}{3}}\phi	\\
\mu&\equiv&\frac{m}{H_0}\,.\label{mu}
\end{eqnarray}
With these definitions, the Friedmann equation can now be written
\begin{equation}
\left(\frac{H}{H_0}\right)^2=\frac{\frac{1}{2}\mu^2\Phi(N)^2+\Omega_{b0}e^{-3N}+\Omega_{\Lambda0}}{1-\frac{1}{2}\Phi'(N)^2}\,.
\label{eqH2}
\end{equation}
With this expression we may find the motion equation as
\begin{equation}
\Phi''+\frac{1}{2E^2}\left[3\mu^2\Phi^2\Phi'+\left(3\Omega_{b0}e^{-3N}+ 6\Omega_{\Lambda0}\right)\Phi'+2\mu^2\Phi\right]=0\,,
\end{equation}
where $E^2\equiv\left(\frac{H}{H_0}\right)^2$ is given by (\ref{eqH2}). This equation can be solved to find the free scalar field evolution. Furthermore, we have the dark matter density parameter
\begin{equation}
\Omega_\phi=\frac{8\pi G\rho_\phi}{3H^2}=\frac{\Phi'^2}{2}+\frac{\mu^2}{2E^2}\Phi^2\,.
\end{equation}
Thus:
\begin{equation}
\Omega_{\phi0}=\frac{\Phi_0'^2}{2}+\frac{\mu^2\Phi^2_0}{2}\,.
\end{equation}

%%%%%%%%%%%%%%%%%%%%%%%%%%%%%%%%%%%%%%%%%%%%%%%%%%%%%%%%%%%%%%%%%%%%%%%%%%

\begin{acknowledgments}
SHP is grateful to CNPq for the financial support, grants No. 304297/2015-1, and to UNESP - Campus Experimental de Itapeva-SP, for the hospitality. FAO was supported by CNPq Brazil through a fellowship within the programme Science without Borders. JLGM is grateful to Prof. C. A. O. Matos for financial support, PAI Carl\~ao grant.
\end{acknowledgments}
%%%%%%%%%%%%%%%%%%%%%%%%%%%%%%%%%%%%%%%%%%%%%%%%%%%%%%%%%%%%%%%%%%%%%%%%%%

\end{document}